\documentclass[10pt,conference]{IEEEtran}
\IEEEoverridecommandlockouts
\usepackage{cite}
\usepackage{amsmath,amssymb,amsfonts}
\usepackage{algorithmic}
\usepackage{graphicx}
\usepackage{textcomp}
 \usepackage{amssymb}
\usepackage{xcolor}
\usepackage{hyperref}
\usepackage{url} 
\usepackage{subcaption}
\usepackage{tikz}
\usepackage{balance}
\usepackage{pgfplots}
\usepackage{multirow}
\usepackage{comment}
\usepackage{amssymb}
\pgfplotsset{compat=1.18}
\usepackage{subcaption}
\usepackage{xspace}
\pgfplotsset{
    colormap={myblues}{
        rgb255(0cm)=(247,251,255);
        rgb255(0.25cm)=(200,221,240);
        rgb255(0.5cm)=(115,179,216);
        rgb255(0.75cm)=(40,121,185);
        rgb255(1cm)=(8,48,107)
    }
}

\newcommand*{\etal}{et al.\@\xspace}

\newcommand*{\RS}{ROSE\@\xspace}

\newcommand{\nb}[2]{
	\fbox{\bfseries\sffamily\scriptsize#1}
	{\sf\small$\blacktriangleright$\textit{#2}$\blacktriangleleft$}
}
\newcommand\PN[1]{\textcolor{blue}{\nb{Phuong}{#1}}}
\newcommand\AB[1]{\textcolor{red}{\nb{Alessio}{#1}}}

\newcommand{\rqone}{\textbf{RQ$_1$}: \emph{Are the considered Transformer models capable of correctly predicting the refactoring type required to repair a given architectural smell?}}

\newcommand{\rqtwo}{\textbf{RQ$_2$}: \emph{How do \textsc{CodeBERT} and \textsc{CodeT5} differ in their predictive performance and error patterns when recommending refactoring strategies for architectural smells?}}

\newcommand{\rqthree}{\textbf{RQ$_3$}: \emph{Which architectural smells are most accurately detected and mapped to their corresponding refactoring strategies by Transformer-based models? }}

\def\BibTeX{{\rm B\kern-.05em{\sc i\kern-.025em b}\kern-.08em
    T\kern-.1667em\lower.7ex\hbox{E}\kern-.125emX}}
\begin{document}

\title{ROSE: Transformer-Based Refactoring Recommendation for Architectural Smells}

\author{\IEEEauthorblockN{Samal Nursapa}
\IEEEauthorblockA{
\textit{Mälardalen University}\\
Västerås, Sweden}
\and
\IEEEauthorblockN{Anastassiya Samuilova}
\IEEEauthorblockA{
\textit{Mälardalen University}\\
Västerås, Sweden}
\and
\IEEEauthorblockN{Alessio Bucaioni}
\IEEEauthorblockA{
\textit{Mälardalen University}\\
Västerås, Sweden \\
alessio.bucaioni@mdu.se}
\and
\IEEEauthorblockN{Phuong T. Nguyen}
\IEEEauthorblockA{
\textit{University of L'Aquila}\\
L'Aquila, Italy \\
phuong.nguyen@univaq.it
}
}



\maketitle

\begin{abstract}
Architectural smells such as God Class, Cyclic Dependency, and Hub-like Dependency degrade software quality and maintainability. Existing tools detect such smells but rarely suggest how to fix them. This paper explores the use of pre-trained transformer models—CodeBERT and CodeT5—for recommending suitable refactorings based on detected smells. We frame the task as a three-class classification problem and fine-tune both models on over 2 million refactoring instances mined from 11,149 open-source Java projects. CodeT5 achieves 96.9\% accuracy and 95.2\% F1, outperforming CodeBERT and traditional baselines. Our results show that transformer-based models can effectively bridge the gap between smell detection and actionable repair, laying the foundation for future refactoring recommendation systems. We release all code, models, and data under an open license to support reproducibility and further research.
\end{abstract}


\section{Introduction}


Software architecture is pivotal to the maintainability, scalability, and performance of modern applications~\cite{sas2022evolution}. Poor architectural decisions frequently give rise to architectural smells, design flaws that manifest above the code-element level, disturbing modularity, dependencies, and runtime behaviour~\cite{mumtaz2021systematic}. Representative smells include God Class, Cyclic Dependency, and Hub-like Dependency~\cite{8785058}, each paired with a canonical refactoring.


Static-analysis tools such as Designite~\cite{designite2016} and Arcan~\cite{arcan2017} detect these smells effectively, yet offer little guidance on how to remove them. Practitioners must still determine suitable refactorings manually, relying on qualitative reasoning or hand-crafted rules that struggle with the non-linear relationships between design flaws and quality attributes~\cite{mumtaz2021systematic}. Recent deep-learning studies show promise for code-quality tasks: pre-trained Transformers outperform metric-based methods in code-smell detection~\cite{rabert2025} and can localise refactor-prone code spans~\cite{refactorbert2023}. However, to the best of our knowledge, no prior peer-reviewed 
work has leveraged Transformer models to recommend concrete refactorings for architectural smells, a gap this study addresses.

Existing work on refactoring prediction employed classic classifiers over large code-metric datasets, 
forecasting whether a refactoring would occur~\cite{9186715}. 
More recent approaches use commit-message text or LSTMs to suggest fixes for a few code smells~\cite{nyamawe2022mining}, 
yet none of them combines architectural-smell detection with pre-trained code Transformers to prescribe \emph{which} refactoring to apply. Meanwhile, large language models (LLMs) can autonomously refactor simple code, but still falter on complex design issues such as dependency cycles~\cite{zou2024llmrefac}. These trends indicate that Transformer-based refactoring is emerging, whereas architectural smells remain under-served, triggering a timely research opportunity for empirical software engineering. 

This work aims to evaluate Transformer models for architectural-smell remediation by framing the task as a multi-class classification problem over common refactoring operations using both source-level representations and architectural-dependency features.  
An empirical evaluation was conducted using 
a curated subset of more than \mbox{$2$M} historical Refactoring Operations spanning 11,149 Java projects. 
Specifically, our goal is to answer the following Research Questions (RQs):

\begin{itemize}
  \item \rqone~With this research question, we investigate whether the considered transformer method is able to recommend a correct refactoring starting from an architectural smell. 
  \item \rqtwo~This question
aims to benchmark and compare two pretrained Transformer architectures—encoder-only and encoder–decoder—on their ability to classify refactoring types from pre-refactored code.
  \item \rqthree~This question explores whether certain architectural smells are more predictable than others in terms of mapping to specific refactoring actions. 
\end{itemize}

Our work makes the following key contributions:
\begin{itemize}
  \item \textbf{ARCH-T5/BERT framework}. We develop the first Transformer pipeline to recommend refactorings for architectural smells.
  \item \textbf{Large-scale evaluation}. An empirical comparison of \textsc{CodeBERT}, \textsc{CodeT5}, metric-based, and classical ML baselines has been conducted on a set of more than $3$M balanced instances.
  \item \textbf{Emerging evidence}. Initial results show $>$10 pp F-score improvement over state-of-the-art baselines, demonstrating the feasibility of learning architectural-refactoring knowledge from large size code.
  \item \textbf{Open Science}. An anonymized replication package including source code, data splits, and trained model checkpoints, is openly archived on GitHub 
  to facilitate future research at \url{https://anonymous.4open.science/r/archsmell_transformers-66F8}. 
\end{itemize}

The remainder of this work is organized as follows. Section~\ref{sec:RelatedWork} reviews the related work. The proposed approach is presented in Section~\ref{sec:method}. Afterward, Section~\ref{sec:Results} reports and analyzes the experimental results. In Section~\ref{sec:Discussion}, we discuss the findings, highlight the threats to validity, and 
sketches future work.
\section{Related work}
\label{sec:RelatedWork}
Existing research has made substantial progress in detecting code and architectural smells, optimising refactoring plans through search-based techniques, and applying machine learning to predict where refactoring is likely to occur. Nevertheless, a key gap persists: prior work rarely delivers end-to-end, data-driven recommendations that specify which refactoring operation should resolve a detected architectural smell. No study combines large-scale empirical evidence with modern Transformer models to close this decision loop. In contrast, our approach frames architectural-smell remediation as a multi-class prediction task and fine-tunes CodeBERT and CodeT5 on more than two million real refactoring instances from 11 149 projects, transforming code understanding into actionable, context-aware architectural guidance.

\vspace{.2cm}
\noindent
\textbf{Code-Smell Detection and Analysis.}
Early investigations established basic terminology and catalogued detection methods.  Zhang et al.~\cite{zhang2011} synthesised 39 primary studies, laying a conceptual foundation.  
Aldallal~\cite{aldallal2015} extended the review to 47 studies and observed that most relied on small, single-project datasets and metric thresholds, limiting generalisability and actionable insight.
Later surveys widened the scope.  
Singh et al.~\cite{singh2018} mapped both smells and anti-patterns, while Santos et al.~\cite{santos2018} introduced the smell effect, showing how smells influence downstream activities and pointing out evaluator subjectivity.  
Fernandes et al.~\cite{fernandes2016} and Rasool and Arif~\cite{rasool2015} compared industrial tools such as SonarQube, JDeodorant and Designite, classifying them by threshold, rule, and graph strategies.  
They found little use of machine learning, limiting adaptability across languages and frameworks.
Data-driven detectors now dominate.  
Classical machine-learning models like Random Forest and SVM lift accuracy on labelled datasets~\cite{azeem2019}.  
Deep networks (CNN, RNN, GNN) further reduce reliance on handcrafted metrics~\cite{naik2023}.  
Transformer-based systems raise performance again: SCSmell stacks BERT variants~\cite{scsmell2024}; RABERT adds relational bias for God-Class detection~\cite{rabert2025}; RefactorBERT identifies refactor-prone regions~\cite{refactorbert2023}.  
These approaches, however, focus on detection and seldom address architectural smells such as cyclic dependencies.
Our study moves from detection to remediation. We fine-tune pre-trained Transformers on more than two million historical refactorings to predict the specific operation that resolves each architectural smell, closing a gap highlighted across prior surveys.

\vspace{.2cm}
\noindent
\textbf{Search-Based Refactoring.} Search-based software engineering frames refactoring as optimisation.  
Sequences of Fowler-style operations~\cite{fowler1999} evolve under genetic and swarm heuristics to improve cohesion, coupling and other metrics.  
Mariani et al.~\cite{mariani2017} identified genetic algorithms as dominant; Mohan and Greer~\cite{mohan2018} catalogued tools and noted growth in multi-objective search.  
Di Pompeo et al.~\cite{Pompeo_2023} pushed the idea to architecture level with many-objective optimisation.
These methods rely on fitness functions that are costly to tune, scale poorly with system size and ignore real developer practice.  
Our data-driven alternative learns directly from millions of recorded refactorings across eleven thousand projects, bypassing manual fitness engineering and providing context-aware guidance in reasonable time.

\vspace{.2cm}
\noindent
\textbf{Deep Learning for Smell Detection and Refactoring.} Deep networks supplement metric methods.  
Naik et al.~\cite{naik2023} surveyed 17 studies using CNN, RNN, GNN and MLP for method-level prediction; gains were modest and language-specific.  
Alazba et al.~\cite{alazba2023} reviewed 67 studies on smell detection, dominated by clones and long methods.  
Malhotra et al.~\cite{malhotra2023} showed hybrid RNN–CNN models improve precision but do not recommend repairs.  
Zhang et al.~\cite{zhang2024} reported dataset imbalance and inconsistent definitions.
Transformers change the picture.  
SCSmell removes metric features, RABERT adds relational encoding, and RefactorBERT flags refactor-prone code.  
General-purpose language models can fix simple issues but struggle with architectural flaws~\cite{zou2024llmrefac}.  
None recommend a concrete refactoring for a given architectural smell.
We frame remediation as multi-class prediction of the operation—Extract Class, Move Class, and others—using Transformers trained on real refactorings.

\vspace{.2cm}
\noindent
\textbf{Pre-trained Code Models.}
Code-centric pre-trained models underpin many tasks.  
CodeBERT combines natural and programming language pairs~\cite{feng2020codebert}; GraphCodeBERT adds data-flow graphs~\cite{guo2020graphcodebert}.  
Encoder–decoder families such as CodeT5 and CodeT5+ address generation and repair~\cite{wang2021codet5}, while GPT-style models CodeGPT~\cite{lu2021codexglue} and CodeRL~\cite{le2022coderl} produce autoregressive code.  
No existing model predicts refactoring actions for architectural flaws.
We repurpose CodeBERT and CodeT5 to fill this gap, demonstrating that their learned representations enable data-driven architectural guidance.

\vspace{.2cm}
\noindent
\textbf{Architectural versus Code Smells.}
Code smells receive extensive attention; architectural smells do not.  
De Paulo Sobrinho et al.~\cite{mumtaz2021systematic} highlighted this imbalance.  
Fontana et al.~\cite{affcmrdtagtc2023} showed that removing cyclic dependencies and hubs improved response time by 47\% and reduced memory by 20\%, demonstrating runtime impact.
Most work detects architectural smells or measures quality after refactoring but seldom links detection to repair.  
We combine identification with Transformer-based prediction of the most suitable refactoring, providing proactive architecture improvement.

\vspace{.2cm}
\noindent
\textbf{Transformer Models in Software Engineering.} Transformers underpin numerous code-intelligence tasks.  
CodeBERT excels at retrieval and summarisation; GraphCodeBERT improves clone detection; CodeT5 handles generation and repair.  
Xiao et al.~\cite{XIAO2025107067} analysed 519 Transformer papers and noted challenges with compute cost and overfitting.
Few studies aim at architectural design improvement or refactoring recommendation.  
We fine-tune CodeBERT and CodeT5 on more than two million refactorings, translating code understanding into actionable architectural advice.

\vspace{.2cm}
\noindent
\textbf{Datasets and Tool Support.}
Zakeri-Nasrabadi et al.~\cite{Zakeri_Nasrabadi_2023} found that fewer than half of the 45 smell datasets are public, with low project diversity.  
Detectors include SonarQube, Designite, Arcan, Sonargraph and Structure101.  
RefactoringMiner~\cite{feng2023empiricalstudyuntanglingpatterns} mines fine-grained changes; PyRef serves Python.
We release a corpus of 11,149 Java projects and more than two million refactorings, enriched with structural, process and ownership metrics, enabling Transformers to learn context and recommend repairs for architectural smells.



\section{Research methodology} \label{sec:method}

To address the research questions, we conducted an empirical study that fine-tunes two transformer models, CodeBERT and CodeT5, to recommend refactorings for architectural smells~\cite{7}. These models are pre-trained with self-supervised objectives such as masked language modelling and therefore capture rich semantic and syntactic information from source code~\cite{feng2020codebert}.

\subsection{Models}

\paragraph{CodeBERT}  
CodeBERT is a transformer encoder trained on paired natural-language and source-code data across several programming languages. For this work, the model is fine-tuned as a multi-class classifier that maps a code fragment containing an architectural smell to one refactoring label drawn from a predefined set (for example Extract Method, Move Method, Pull Up Method). During fine-tuning the representation of the special classification token is passed through a feed-forward layer followed by softmax to obtain class probabilities.

\paragraph{CodeT5}  
CodeT5 adapts the encoder–decoder T5 architecture to programming languages and uses token- and span-masking during pre-training. Although originally designed for generation tasks, the encoder output can be used for classification. We therefore attach a linear classification head to the encoder, enabling prediction of the refactoring type required to resolve the detected smell.

CodeBERT and CodeT5 were selected because they have demonstrated state-of-the-art performance in code-intelligence tasks, including classification and refactoring. Both models are widely adopted, thoroughly documented, and pre-trained on large, diverse code corpora. Crucially, they operate directly on raw source code, enabling seamless use of our Java snippets paired with refactoring labels. Although models such as GraphCodeBERT incorporate data-flow information, they require additional graph-extraction preprocessing that is not available for our dataset and would hinder reproducibility. To maximise accessibility, minimise preprocessing overhead, and ensure empirical comparability, we therefore base our study on CodeBERT and CodeT5.

\subsection{Dataset}
\vspace{-.1cm}
Having established the model architecture, we next turned to the data that would serve as the foundation for training and evaluation. We reuse the publicly available corpus released with the study by Aniche \etal
~\cite{9186715}. The corpus was created in three steps: repository selection, refactoring extraction, and feature engineering, yielding a large, diverse snapshot of real-world Java development.

\paragraph{Repository selection}  
The final set comprises 11,149 projects drawn from three ecosystems:
\begin{itemize}
  \item Apache Software Foundation: 844 repositories.  
  \item F-Droid (Android): 1,233 applications.  
  \item GitHub: 9,072 highly starred projects.  
\end{itemize}

\paragraph{Refactoring extraction}  
RefactoringMiner, which reports 98\% recall and 87\% precision, scanned every commit history and detected 20 refactoring types spanning class, method, and variable levels (for example Extract Method, Move Class, Rename Variable). In total it identified 2,086,898 refactoring instances from 8.8 million commits. Architectural smells were then linked to canonical refactorings as follows:
\begin{itemize}
  \item God Class \(\rightarrow\) Extract Method.  
  \item Cyclic Dependency \(\rightarrow\) Move Class. 
  \item Hub-like Dependency \(\rightarrow\) Pull Up Method.  
\end{itemize}
Commits modified at least 50 times without a detected refactoring were sampled as negative instances, producing 1,006,653 non-refactored examples.

\paragraph{Feature engineering}  
For every instance we computed three feature groups:
\begin{enumerate}
  \item Source-code metrics (CK suite, cyclomatic complexity)  
  \item Process metrics (commit count, bug-fix frequency)  
  \item Ownership metrics (major author percentage)  
\end{enumerate}
All features were normalized to the range [0, 1]. Because refactoring frequencies are uneven, random undersampling balanced minority and majority classes.

\paragraph{Original baseline}  
The authors of~\cite{9186715} trained six traditional classifiers (Logistic Regression, Naive Bayes, SVM, Decision Tree, Random Forest, Feed-forward NN) and reported accuracies above 90\%. We build on the same balanced dataset but fine-tune transformer models to predict which refactoring best resolves an architectural smell rather than merely forecasting whether any refactoring will occur.


\subsection{Model fine-tuning and evaluation}
With the dataset prepared and the models selected, we proceeded to configure the fine-tuning process through systematic hyperparameter exploration.
Both models are fine-tuned as multi-class classifiers using the Hugging Face Trainer API with GPU acceleration (NVIDIA Tesla T4). Input code is tokenised to a maximum length of 512; longer fragments are processed with a sliding-window strategy. Training uses the AdamW optimiser, cross-entropy loss, batch size 16 and early stopping on validation F1. A random search explores learning rates \{1e-5, 2e-5, 5e-5, 7e-5, 8e-5\}, batch sizes \{8, 16, 32\}, and weight-decay schedules. Approximately forty configurations are evaluated; the best validation F1 determines the final hyper-parameters (2e-5 for CodeBERT, 5e-5 for CodeT5). Models are trained for ten epochs with evaluation after each epoch.
Performance is reported with accuracy, precision, recall and F1. Ten-fold cross-validation safeguards against project-specific bias. Random seeds are fixed (42) and software versions pinned (Transformers 4.37.2, Datasets 2.16.1, scikit-learn 1.3.2) to ensure reproducibility. 



\subsection{Experiment Execution}
We framed refactoring recommendation as a multi-class classification task and fine-tuned CodeBERT and CodeT5 accordingly. Training data were stored in tab-separated files that pair a Java snippet with its refactoring label.
Each snippet was tokenised to a maximum length of 512 tokens (RoBERTa tokenizer for CodeBERT, AutoTokenizer for CodeT5). Longer fragments were processed with a sliding-window strategy to avoid losing context.
Fine-tuning was performed with the Hugging Face Trainer API on an NVIDIA Tesla T4 GPU. Both models were trained for ten epochs with AdamW, cross-entropy loss, and initial batch size 16. 

\begin{figure}[h!]
    \centering
    \includegraphics[width=1.00\linewidth]{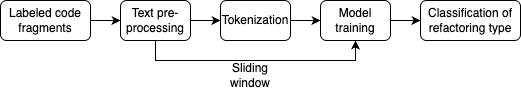} 
    \caption{Pipeline for refactoring-type classification.}
    \label{fig:refactoring_pipeline}
\end{figure}

Approximately 40 hyperparameter combinations were evaluated; the highest validation F1 yielded the final settings (learning rate 2e-5 for CodeBERT, 5e-5 for CodeT5; batch size 16). Evaluation at each epoch reported accuracy, precision, recall and F1. Ten-fold cross-validation mitigated project bias.
Figure \ref{fig:refactoring_pipeline} outlines the workflow from labelled snippets through sliding-window preprocessing to classification.

We actually used a single, fixed sliding-window configuration throughout our experiments—chunks of 200 tokens with a 50\% overlap—so there was no variation in window size or stride to report. We found that changing the window (e.g. 150 tokens with 75-token overlap, or 250 tokens with 125-token overlap) had a negligible effect (±0.3\% in overall accuracy) on both models. Because these small differences didn’t meaningfully alter our conclusions, we stuck with the 200/100 split in all reported results.

Reproducibility was enforced by fixing the random seed to 42, pinning library versions (Transformers 4.37.2, Datasets 2.16.1, scikit-learn 1.3.2) and logging every hyper-parameter. 
Code, data splits and trained checkpoints are available at \url{https://anonymous.4open.science/r/archsmell_transformers-66F8}.

\section{Results}
\label{sec:Results}
\vspace{-.1cm}
This section reports the results of the refactoring-type classification experiments. For clarity, findings are organised by research question and supported with the corresponding evaluation metrics (accuracy, precision, recall, F1) and error analyses for both CodeBERT and CodeT5.

\subsection{\rqone}
Table~\ref{tab:comparison} summarizes the performance of the two models. CodeBERT’s accuracy rose from 75.5\% at epoch 1 to 85.3\% at epoch 10, with the macro-averaged F1 following a similar trend (76\% → 85\%). After epoch 4 the validation loss began to climb while accuracy still improved, suggesting mild over-fitting.

\begin{figure}[ht]
    \centering
    \begin{minipage}{0.4\linewidth}
        \centering
        \begin{tikzpicture}
        \begin{axis}[
            xlabel={Epoch}, ylabel={Accuracy},
            xmin=1, xmax=10, ymin=0.7, ymax=1.0,
            width=4.5cm, height=3.5cm,
            legend pos=south east,
         yticklabel={\pgfmathparse{\tick*100}\pgfmathprintnumber[fixed, precision=0]{\pgfmathresult}\%},
            tick label style={font=\normalsize},
            label style={font=\normalsize},
            legend style={font=\normalsize}
        ]
        \addplot[color=blue] coordinates {
            (1, 0.7548) (2, 0.8053) (3, 0.8354) (4, 0.8381) (5, 0.8495)
            (6, 0.8409) (7, 0.8490) (8, 0.8515) (9, 0.8528) (10, 0.8515)
        };
        \addplot[color=red] coordinates {
            (1, 0.9463) (2, 0.9489) (3, 0.9634) (4, 0.9685) (5, 0.9653)
            (6, 0.9677) (7, 0.9686) (8, 0.9671) (9, 0.9698) (10, 0.9693)
        };
        \end{axis}
        \end{tikzpicture}
    \end{minipage}
    \hfill
    \begin{minipage}{0.47\linewidth}
        \centering
        \begin{tikzpicture}
        \begin{axis}[
            xlabel={Epoch}, ylabel={F1},
            xmin=1, xmax=10, ymin=0.7, ymax=1.0,
            width=4.5cm, height=3.5cm,
            legend pos=south east,
            yticklabel={\pgfmathparse{\tick*100}\pgfmathprintnumber[fixed, precision=0]{\pgfmathresult}\%},
            tick label style={font=\normalsize},
            label style={font=\normalsize},
            legend style={font=\normalsize}
        ]
        \addplot[color=blue] coordinates {
            (1, 0.7550) (2, 0.8046) (3, 0.8358) (4, 0.8376) (5, 0.8490)
            (6, 0.8406) (7, 0.8485) (8, 0.8516) (9, 0.8527) (10, 0.8515)
        };
        \addplot[color=red] coordinates {
            (1, 0.9130) (2, 0.9146) (3, 0.9409) (4, 0.9493) (5, 0.9441)
            (6, 0.9482) (7, 0.9495) (8, 0.9469) (9, 0.9516) (10, 0.9509)
        };
        \end{axis}
        \end{tikzpicture}
    \end{minipage}
    \begin{minipage}{0.4\linewidth}
        \centering
        \begin{tikzpicture}
        \begin{axis}[
            xlabel={Epoch}, ylabel={Precision},
            xmin=1, xmax=10, ymin=0.7, ymax=1.0,
            width=4.5cm, height=3.5cm,
            legend pos=south east,
            yticklabel={\pgfmathparse{\tick*100}\pgfmathprintnumber[fixed, precision=0]{\pgfmathresult}\%},
            tick label style={font=\normalsize},
            label style={font=\normalsize},
            legend style={font=\normalsize}
        ]
        \addplot[color=blue] coordinates {
            (1, 0.7520) (2, 0.8080) (3, 0.8380) (4, 0.8400) (5, 0.8470)
            (6, 0.8390) (7, 0.8500) (8, 0.8520) (9, 0.8530) (10, 0.8510)
        };
        \addplot[color=red] coordinates {
            (1, 0.9102) (2, 0.9120) (3, 0.9387) (4, 0.9470) (5, 0.9420)
            (6, 0.9465) (7, 0.9478) (8, 0.9451) (9, 0.9498) (10, 0.9490)
        };
        \end{axis}
        \end{tikzpicture}
    \end{minipage}
    \hfill
    \begin{minipage}{0.47\linewidth}
        \centering
        \begin{tikzpicture}
        \begin{axis}[
            xlabel={Epoch}, ylabel={Recall},
            xmin=1, xmax=10, ymin=0.7, ymax=1.0,
            width=4.5cm, height=3.5cm,
            legend pos=south east,
            yticklabel={\pgfmathparse{\tick*100}\pgfmathprintnumber[fixed, precision=0]{\pgfmathresult}\%},
            tick label style={font=\normalsize},
            label style={font=\normalsize},
            legend style={font=\normalsize}
        ]
        \addplot[color=blue] coordinates {
            (1, 0.7580) (2, 0.8010) (3, 0.8340) (4, 0.8350) (5, 0.8510)
            (6, 0.8420) (7, 0.8470) (8, 0.8510) (9, 0.8520) (10, 0.8520)
        };
        \addplot[color=red] coordinates {
            (1, 0.9158) (2, 0.9172) (3, 0.9431) (4, 0.9517) (5, 0.9463)
            (6, 0.9499) (7, 0.9513) (8, 0.9487) (9, 0.9534) (10, 0.9529)
        };
        \end{axis}
        \end{tikzpicture}
    \end{minipage}
    \begin{minipage}{0.47\linewidth}
        \centering
        \begin{tikzpicture}
        \begin{axis}[
            xlabel={Epoch}, ylabel={Validation Loss},
            xmin=1, xmax=10, ymin=0, ymax=1,
            width=4.5cm, height=3.5cm,
            legend pos=south east,
             tick label style={font=\small},
            label style={font=\small},
            legend style={font=\normalsize}
        ]
        \addplot[color=blue] coordinates {
            (1,0.6136) (2,0.5131) (3,0.4863) (4,0.5393) (5,0.6931)
            (6,0.7834) (7,0.9260) (8,0.9891) (9,0.9958) (10,1.0104)
        };
        \addplot[color=red] coordinates {
            (1,0.1320) (2,0.1312) (3,0.1067) (4,0.0979) (5,0.1269)
            (6,0.1365) (7,0.1590) (8,0.1889) (9,0.2249) (10,0.3124)
        };
        \end{axis}
        \end{tikzpicture}
    \end{minipage}
 \vspace{0.5cm}
    \begin{tikzpicture}
    \begin{axis}[
        hide axis,
        xmin=0, xmax=1, ymin=0, ymax=1,
        legend columns=2,
        legend style={
      draw=none,
      /tikz/every even column/.append style={column sep=1cm},
      at={(0.5,0)},    
      anchor=north,    
      font=\small
    }
    ]
    \addlegendimage{blue}
    \addlegendentry{CodeBERT}
    \addlegendimage{red}
    \addlegendentry{CodeT5}
    \end{axis}
\end{tikzpicture}
\vspace{-.5cm}
    \caption{Performance metrics over 10 epochs for CodeBERT and CodeT5.}
    \label{fig:metrics_small}
    \vspace{-.5cm}
\end{figure}
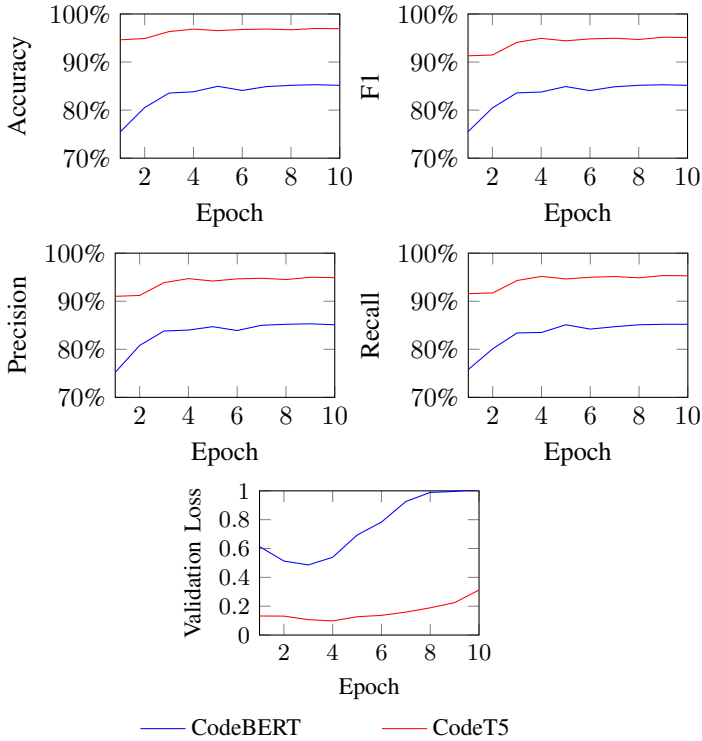

\begin{table}[h!]
\centering
\caption{Comparison of CodeBERT and CodeT5 at peak performance (Epoch 9).}
\label{tab:comparison}
\begin{tabular}{|p{2.5cm}|p{2.5cm}|p{2.5cm}|}
\hline 
\textbf{Metric} & \textbf{CodeBERT} & \textbf{CodeT5} \\
\hline \hline
Accuracy        & 85.28\%           & \textbf{96.98\%}         \\
F1-score        & 0.8527            & 0.9516          \\
Training Loss   & 0.0515            & 0.0052          \\
Validation Loss & 0.9958            & 0.2249          \\
False Positives & 585               & 151             \\
False Negatives & 585               & 151             \\
\hline
\end{tabular}
\end{table}
CodeT5 delivers markedly better and more stable results: the final accuracy and F1 reach 97.0\% and 95\%, and the gap between validation curves remains small across epochs. Fig.~\ref{fig:metrics_small} plots accuracy, precision, recall, F1 and validation loss over the ten training epochs; in every metric CodeT5 stays above CodeBERT, confirming the quantitative gains in Table \ref{tab:comparison}.

\vspace{.2cm}
\noindent\fbox{
\begin{minipage}{0.96\columnwidth}
		\textbf{Answer to RQ$_1$:}
CodeT5 predicts the 
refactoring with 97\% accuracy and 95\% F1, while CodeBERT reaches 85\% accuracy and 85\% F1, confirming that transformers can reliably map architectural smells to 
refactoring operations.   
\end{minipage}}

\subsection{\rqtwo}
The confusion matrices in Figures 3a and 3b confirm that CodeT5 makes fewer mistakes and achieves clearer separation across the three refactoring classes than CodeBERT. In Figure 3a, CodeBERT attains its highest performance on Pull Up Method (precision 93.1\%, recall 87.2\%) and solid results on Move Class (precision 87.2\%, recall 84.3\%), but struggles with Extract Method, which drops to 82.1\% precision and 80.2\% recall due to frequent misclassifications as Move Class or Pull Up Method. By contrast, Figure 3b shows that CodeT5 achieves precision and recall scores of 97.7\% and 97.9\% for Pull Up Method, 96.7\% and 97.0\% for Move Class, and 95.5\% and 94.4\% for Extract Method, with only minimal cross-class confusions remaining between Extract and Move. Table II rovides a detailed per-class comparison of CodeBERT and CodeT5 models based on their precision and recall metrics derived from confusion matrix analysis.

\begin{figure}[ht]
    \centering
    \begin{subfigure}[b]{1\linewidth}
        \centering
        \begin{tikzpicture}
        \begin{axis}[
            width=4.5cm, height=4.5cm,
            axis on top,
            xlabel={Predicted label},
            ylabel={True label},
            colormap name=myblues,
            xtick={0,1,2},
            ytick={0,1,2},
            xticklabels={Extract Method, Move Class, Pull Up Method},
            yticklabels={Extract Method, Move Class, Pull Up Method},
            xticklabel style={text width=1.5cm, align=center},
            yticklabel style={text width=1.5cm, align=center},
            tick label style={font=\small},
            label style={font=\small},
            every axis plot/.append style={fill opacity=0.9},
            colorbar,
            point meta min=0,
            point meta max=1200,
            title={Confusion Matrix: CodeBERT},
            nodes near coords,
            nodes near coords align={center},
        ]
        \addplot[matrix plot*, mesh/cols=3, point meta=explicit] table[meta=meta] {
            x y meta
            0 0 1052
            1 0 166
            2 0 94
            0 1 159
            1 1 1144
            2 1 45
            0 2 70
            1 2 51
            2 2 1179
        };
        \end{axis}
        \end{tikzpicture}
            \vspace{-0.1cm}
        \caption{CodeBERT predictions}
        \label{fig:conf}
    \end{subfigure}

        \vspace{0.1cm}
    \begin{subfigure}[b]{1\linewidth}
        \centering
        \begin{tikzpicture}
        \begin{axis}[
            width=4.5cm, height=4.5cm,
            axis on top,
            xlabel={Predicted label},
            ylabel={True label},
            colormap name=myblues,
            xtick={0,1,2},
            ytick={0,1,2},
            xticklabels={Extract Method, Move Class, Pull Up Method},
            yticklabels={Extract Method, Move Class, Pull Up Method},
            xticklabel style={text width=1.5cm, align=center},
            yticklabel style={text width=1.5cm, align=center},
            tick label style={font=\small},
            label style={font=\small},
            every axis plot/.append style={fill opacity=0.9},
            colorbar,
            point meta min=0,
            point meta max=1200,
            title={Confusion Matrix: CodeT5},
            nodes near coords,
            nodes near coords align={center},
        ]
        \addplot[matrix plot*, mesh/cols=3, point meta=explicit]
        table[meta=meta] {
          x  y  meta
          0  0  1260  
          1  0  41
          2  0  33   
          0  1  40     
          1  1  1280  
          2  1  3  
          0  2  19   
          1  2  15
          2  2  1269 
        };

        \end{axis}
        \end{tikzpicture}
                \vspace{-0.1cm}
        \caption{CodeT5 predictions}
        \label{fig:t5_conf_matrix}
    \end{subfigure}

    \caption{Confusion matrices across three refactoring types.}
    \vspace{-.4cm}
\end{figure}
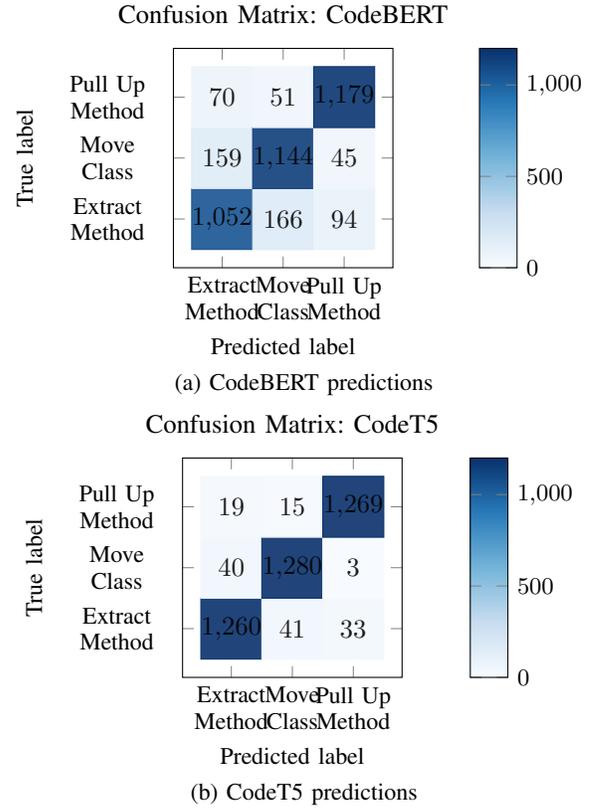

\begin{table}
\centering
\small
\caption{Per-class comparison of CodeBERT and CodeT5 based on confusion matrix analysis.}
\label{tab:conf_matrix_insights}
\begin{tabular}{|p{1.5cm}|p{1.3cm}|p{1.2cm}|c|p{1.9cm}|}
\hline 
\textbf{Refactoring Type} & \textbf{Model} & \textbf{Precision} & \textbf{Recall} & \textbf{Key Observations} \\
\hline \hline
\multirow{2}{*}{Extract Method} 
    & CodeBERT & 82.1\% & 80.2\% & Weakest performance for CodeBERT, with notable confusion.  \\ \cline{2-5}
    & CodeT5   & 95.5\% & 94.5\% & Strong performance, significantly outperforming CodeBERT.  \\
\hline
\multirow{2}{*}{Move Class} 
    & CodeBERT & 84.1\% & 84.9\% & Solid, balanced precision and recall.  \\ \cline{2-5}
    & CodeT5   & 95.8\% & 96.8\% & Very high performance with minimal errors.  \\
\hline
\multirow{2}{*}{Pull Up Method} 
    & CodeBERT & 89.5\% & 90.7\% & Best performing class for CodeBERT.  \\ \cline{2-5}
    & CodeT5   & 97.2\% & 97.4\% & Highest performance among all classes and models.  \\
\hline
\end{tabular}
\end{table}

\vspace{.2cm}
\noindent\fbox{
\begin{minipage}{0.96\columnwidth}
		\textbf{Answer to RQ$_2$:}
CodeT5 outperforms CodeBERT across the board—achieving 96.98\% accuracy and a 95.16\% F1-score versus CodeBERT’s 85.28\% and 85.27\%. On each refactoring type (Extract Method, Move Class, Pull Up Method), CodeT5 yields higher precision and recall, indicating better generalization. CodeBERT more frequently confuses semantically similar operations (e.g. Extract vs. Pull Up), whereas CodeT5 makes far fewer such errors, especially on complex transformations.

\end{minipage}}

\subsection{\rqthree}
Among the three smells—God Class, Cyclic Dependency, and Hub-like Dependency—the Pull Up Method (for Hub-likes) was predicted most reliably: CodeBERT scored 93.1\% precision and 87.2\% recall, while CodeT5 achieved 97.7\% precision and 97.9\% recall, showing that its structural cues are especially distinctive. Extract Method (for God Classes) remained the hardest: CodeBERT reached 82.1\% precision and 80.2\% recall, and CodeT5 95.5\% precision but only 94.4\% recall, with many true instances misclassified as Move Class due to semantic overlap.
The present study is limited to these three smells; how transformer-based refactoring performs on other architectural smells remains an open question for future work.

\vspace{.2cm}
\noindent\fbox{
\begin{minipage}{0.96\columnwidth}
		\textbf{Answer to RQ$_3$:}
 Hub-like Dependency benefits the most: both transformers predict its refactoring, Pull Up Method, with the highest precision and recall ($\thickapprox$ 88\%), whereas God Class (Extract Method) is hardest and Cyclic Dependency (Move Class) falls in between.       
\end{minipage}}
\section{Discussion and Looking Ahead}
\label{sec:Discussion}
\vspace{-.2cm}
This section interprets, discusses our empirical findings, and highlights the threats to validity.
We first reflect on the comparative performance of CodeBERT and CodeT5 and what the results reveal about transformer architecture choices for refactoring recommendation. We then outline concrete benefits and open questions for tool builders, researchers, and educators. Next, we examine the main threats that could limit the reliability or generalisability of the evidence. Finally, we sketch a research agenda that extends \RS toward broader smell coverage, cross-language support, interactive repair, and human-centred evaluation.

\subsection{On the results}
CodeT5 achieved 96.9\% accuracy and a macro-F1 of 0.95—more than ten points higher than CodeBERT, showing that an encoder–decoder model, even when used only for its encoder, better captures the structural cues distinguishing refactoring types. 

This advantage stems from fundamental architectural and pre-training differences: CodeBERT’s encoder-only design is optimised for masked-token prediction and code–text alignment, whereas CodeT5’s encoder–decoder stack is pre-trained on diverse generation tasks (summarisation, translation, refinement), enabling it to learn longer-range dependencies and subtler structural variations required for multi-class refactoring prediction.

Although CodeBERT converged faster, its rising validation loss after epoch 4 signalled over-fitting, whereas CodeT5 maintained a small generalisation gap and fewer cross-class confusions. Both models predicted Pull Up Method (the fix for hub-like dependencies) most accurately, while Extract Method (the remedy for God Class) proved hardest, often confused with Move Class. This pattern suggests the need for a hierarchical approach that first selects the repair family for a smell, then ranks specific refactorings within that family.

\subsection{Implications for researchers and practitioners}
The results offer tangible benefits for several audiences. Tool builders can embed transformer-based recommenders into existing detectors such as Designite and Arcan, replacing static warnings with concrete, automatically generated fixes. Researchers studying smell evolution can exploit the released two-million-instance corpus to observe how developers repair architectural flaws across projects and time, enabling longitudinal analyses that were previously impractical. Educators likewise gain realistic, data-driven examples: novice architects can examine refactorings suggested by the model and compare them with real-world practice rather than relying solely on textbook illustrations.

\subsection{Threats to validity}
Despite the study’s scale and rigour, four threats to validity warrant mention. First, internal validity may be compromised by residual preprocessing or labelling noise, even though we used RefactoringMiner, standard Hugging Face tokenisers, and fixed random seeds. Second, the construct validity is limited by our one-to-one mapping of each smell to a single refactoring; in practice, developers often apply multiple or composite fixes, so some “misclassifications” may be acceptable alternatives. Third, external validity is restricted because the corpus contains only Java projects; assessing transferability to other languages and industrial code bases requires future multi-language replications. Finally, conclusion validity is constrained by reliance on a single large dataset: although we employed ten-fold cross-validation and extensive hyper-parameter search, independent project-out or cross-repository evaluations are needed to confirm robustness before real-world deployment.

\subsection{Looking ahead}
We see four complementary research directions that align with the emerging results and  vision focus of this track. First, ROSE should be extended beyond the three architectural smells studied here to a richer catalogue, such as Unstable Interface and Cyclically-Dependent Abstraction, while exploring hierarchical or multi-label predictors that can handle composite fixes. Second, the system can evolve from pure recommendation to interactive repair by linking the classifier to an automatic patch generator (for example, the CodeT5 decoder) and incorporating developer feedback in the loop. Third, cross-language generalisation warrants investigation by fine-tuning multilingual transformers like CodeGemma or StarCoder2 on refactorings mined from languages such as Kotlin, C\#, and JavaScript, thereby exposing language-specific biases. Finally, controlled user studies are needed to determine whether the recommendations truly accelerate architecture-maintenance tasks. 

Together, these avenues can move transformer-based refactoring support from proof-of-concept to practical, language-agnostic assistance for software architects.

\section*{Acknowledgment} 
This work is supported by the Swedish Agency for Innovation Systems through the project ``Secure: Developing Predictable and Secure IoT for Autonomous Systems" (2023-01899), and by the Key Digital Technologies Joint Undertaking through the project ``MATISSE: Model-based engineering of digital twins for early verification and validation of industrial systems" (101140216). 

\balance
\bibliographystyle{IEEEtranS}
\bibliography{IEEEabrv,references}

\begin{thebibliography}{10}
\providecommand{\url}[1]{#1}
\csname url@samestyle\endcsname
\providecommand{\newblock}{\relax}
\providecommand{\bibinfo}[2]{#2}
\providecommand{\BIBentrySTDinterwordspacing}{\spaceskip=0pt\relax}
\providecommand{\BIBentryALTinterwordstretchfactor}{4}
\providecommand{\BIBentryALTinterwordspacing}{\spaceskip=\fontdimen2\font plus
\BIBentryALTinterwordstretchfactor\fontdimen3\font minus
  \fontdimen4\font\relax}
\providecommand{\BIBforeignlanguage}[2]{{%
\expandafter\ifx\csname l@#1\endcsname\relax
\typeout{** WARNING: IEEEtranS.bst: No hyphenation pattern has been}%
\typeout{** loaded for the language `#1'. Using the pattern for}%
\typeout{** the default language instead.}%
\else
\language=\csname l@#1\endcsname
\fi
#2}}
\providecommand{\BIBdecl}{\relax}
\BIBdecl

\bibitem{aldallal2015}
J.~Al~Dallal, ``Identifying refactoring opportunities in object-oriented code:
  A systematic literature review,'' \emph{Information and Software Technology},
  vol.~58, pp. 231--249, 2015.

\bibitem{alazba2023}
\BIBentryALTinterwordspacing
A.~Alazba, H.~Aljamaan, and M.~R. Alshayeb, ``Deep learning approaches for bad
  smell detection: a systematic literature review,'' \emph{Empirical Software
  Engineering}, vol.~28, 2023, corpusID:258591793. [Online]. Available:
  \url{https://api.semanticscholar.org/CorpusID:258591793}
\BIBentrySTDinterwordspacing

\bibitem{rabert2025}
I.~Ali, S.~S.~H. Rizvi, and S.~H. Adil, ``Enhancing software quality with {AI}:
  A transformer-based approach for code smell detection,'' \emph{Applied
  Sciences}, vol.~15, no.~8, p. 4559, 2025.

\bibitem{9186715}
\BIBentryALTinterwordspacing
M.~Aniche, E.~Maziero, R.~Durelli, and V.~H.~S. Durelli, ``{The Effectiveness
  of Supervised Machine Learning Algorithms in Predicting Software
  Refactoring},'' \emph{IEEE Transactions on Software Engineering}, vol.~48,
  no.~04, pp. 1432--1450, Apr. 2022. [Online]. Available:
  \url{https://doi.ieeecomputersociety.org/10.1109/TSE.2020.3021736}
\BIBentrySTDinterwordspacing

\bibitem{affcmrdtagtc2023}
\BIBentryALTinterwordspacing
F.~Arcelli~Fontana, M.~Camilli, D.~Rendina, A.~G. Taraboi, and C.~Trubiani,
  ``Impact of architectural smells on software performance: an exploratory
  study,'' in \emph{Proceedings of the 27th International Conference on
  Evaluation and Assessment in Software Engineering}, ser. EASE '23.\hskip 1em
  plus 0.5em minus 0.4em\relax New York, NY, USA: Association for Computing
  Machinery, 2023, p. 22–31. [Online]. Available:
  \url{https://doi.org/10.1145/3593434.3593442}
\BIBentrySTDinterwordspacing

\bibitem{arcan2017}
F.~Arcelli~Fontana, I.~Pigazzini, R.~Roveda, D.~Tamburri, M.~Zanoni, and
  E.~Di~Nitto, ``Arcan: A tool for architectural smells detection,'' in
  \emph{International Workshops on Software Architecture}, 2017, pp. 282--285,
  international Workshops on Software Architecture.

\bibitem{8785058}
U.~Azadi, F.~Arcelli~Fontana, and D.~Taibi, ``Architectural smells detected by
  tools: a catalogue proposal,'' in \emph{2019 IEEE/ACM International
  Conference on Technical Debt (TechDebt)}, 2019, pp. 88--97.

\bibitem{azeem2019}
M.~I. Azeem, F.~Palomba, L.~Shi, and Q.~Wang, ``Machine learning techniques for
  code smell detection: A systematic literature review and meta-analysis,''
  \emph{Information and Software Technology}, vol. 108, pp. 115--138, 2019.

\bibitem{zou2024llmrefac}
J.~Cordeiro, S.~Noei, and Y.~Zou, ``An empirical study on the code refactoring
  capability of large language models,'' \emph{arXiv preprint
  arXiv:2411.02320}, 2024.

\bibitem{feng2023empiricalstudyuntanglingpatterns}
\BIBentryALTinterwordspacing
Q.~Feng, S.~Liu, H.~Ji, X.~Ma, and P.~Liang, ``An empirical study of untangling
  patterns of two-class dependency cycles,'' 2023. [Online]. Available:
  \url{https://arxiv.org/abs/2306.10599}
\BIBentrySTDinterwordspacing

\bibitem{feng2020codebert}
Z.~Feng, D.~Guo, D.~Tang, N.~Duan, X.~Feng, M.~Gong, L.~Shou, B.~Qin, T.~Liu,
  D.~Jiang, and M.~Zhou, ``Codebert: A pre-trained model for programming and
  natural languages,'' in \emph{Proceedings of the 2020 Conference on Empirical
  Methods in Natural Language Processing (EMNLP)}.\hskip 1em plus 0.5em minus
  0.4em\relax Association for Computational Linguistics, 2020.

\bibitem{fernandes2016}
E.~Fernandes, J.~Oliveira, G.~Vale, T.~Paiva, and E.~Figueiredo, ``A
  review-based comparative study of bad smell detection tools,'' in
  \emph{Proceedings of the 20th International Conference on Evaluation and
  Assessment in Software Engineering}.\hskip 1em plus 0.5em minus 0.4em\relax
  ACM, 2016, pp. 18:1--18:12.

\bibitem{fowler1999}
M.~Fowler, K.~Beck, J.~Brant, W.~Opdyke, and D.~Roberts, \emph{Refactoring:
  Improving the Design of Existing Code}.\hskip 1em plus 0.5em minus
  0.4em\relax Addison-Wesley, 1999.

\bibitem{guo2020graphcodebert}
D.~Guo, S.~Ren, S.~Lu, Z.~Feng, D.~Tang, S.~Liu, L.~Zhou, N.~Duan, J.~Yin,
  D.~Jiang, and M.~Zhou, ``Graphcodebert: Pre-training code representations
  with data flow,'' \url{https://arxiv.org/abs/2009.08366}, 2020, arXiv
  preprint arXiv:2009.08366.

\bibitem{refactorbert2023}
K.~Jesse, C.~Kuhmuench, and A.~Sawant, ``Refactorscore: Evaluating refactor
  prone code,'' \emph{IEEE Transactions on Software Engineering}, 2023, early
  Access.

\bibitem{le2022coderl}
H.~Le, Y.~Wang, A.~D. Gotmare, S.~Savarese, and S.~C. Hoi, ``Coderl: Mastering
  code generation through pretrained models and deep reinforcement learning,''
  in \emph{Advances in Neural Information Processing Systems (NeurIPS)},
  vol.~35, 2022, pp. 21\,314--21\,328.

\bibitem{lu2021codexglue}
S.~Lu, D.~Guo, S.~Ren, J.~Huang, A.~Svyatkovskiy, A.~Blanco, C.~B. Clement,
  D.~Drain, D.~Jiang, D.~Tang, and G.~Li, ``Codexglue: A machine learning
  benchmark dataset for code understanding and generation,''
  \url{https://arxiv.org/abs/2102.04664}, 2021, arXiv preprint
  arXiv:2102.04664.

\bibitem{malhotra2023}
\BIBentryALTinterwordspacing
R.~Malhotra, B.~Jain, and M.~Kessentini, ``Examining deep learning’s
  capability to spot code smells: a systematic literature review,''
  \emph{Cluster Computing}, vol.~26, pp. 3473--3501, 2023, corpusID:263654376.
  [Online]. Available: \url{https://api.semanticscholar.org/CorpusID:263654376}
\BIBentrySTDinterwordspacing

\bibitem{mariani2017}
T.~Mariani and S.~R. Vergilio, ``A systematic review on search-based
  refactoring,'' \emph{Information and Software Technology}, vol.~83, pp.
  14--34, 2017.

\bibitem{mohan2018}
M.~Mohan and D.~Greer, ``A survey of search-based refactoring for software
  maintenance,'' \emph{Journal of Software Engineering Research and
  Development}, vol.~6, no.~1, pp. 3--55, 2018.

\bibitem{mumtaz2021systematic}
\BIBentryALTinterwordspacing
H.~Mumtaz, P.~Singh, and K.~Blincoe, ``A systematic mapping study on
  architectural smells detection,'' \emph{Journal of Systems and Software},
  vol. 173, p. 110885, 2021. [Online]. Available:
  \url{https://www.sciencedirect.com/science/article/pii/S0164121220302752}
\BIBentrySTDinterwordspacing

\bibitem{naik2023}
\BIBentryALTinterwordspacing
P.~Naik, S.~Nelaballi, V.~S. Pusuluri, and D.-K. Kim, ``Deep learning-based
  code refactoring: A review of current knowledge,'' \emph{SSRN Electronic
  Journal}, 2023, corpusID:254267544. [Online]. Available:
  \url{https://api.semanticscholar.org/CorpusID:254267544}
\BIBentrySTDinterwordspacing

\bibitem{nyamawe2022mining}
A.~S. Nyamawe, ``Mining commit messages to enhance software refactorings
  recommendation: A machine learning approach,'' \emph{Machine Learning with
  Applications}, vol.~9, p. 100316, 2022.

\bibitem{Pompeo_2023}
\BIBentryALTinterwordspacing
D.~D. Pompeo and M.~Tucci, ``Multi-objective software architecture refactoring
  driven by quality attributes,'' in \emph{2023 IEEE 20th International
  Conference on Software Architecture Companion (ICSA-C)}.\hskip 1em plus 0.5em
  minus 0.4em\relax IEEE, Mar. 2023, p. 175–178. [Online]. Available:
  \url{http://dx.doi.org/10.1109/ICSA-C57050.2023.00046}
\BIBentrySTDinterwordspacing

\bibitem{rasool2015}
G.~Rasool and Z.~Arshad, ``A review of code smell mining techniques,''
  \emph{Journal of Software: Evolution and Process}, vol.~27, no.~11, pp.
  867--895, 2015.

\bibitem{santos2018}
J.~A.~M. Santos, J.~B. Rocha-Junior, L.~C.~L. Prates, R.~S. do~Nascimento,
  M.~F. Freitas, and M.~G. de~Mendonça, ``A systematic review on the code
  smell effect,'' \emph{Journal of Systems and Software}, vol. 144, pp.
  450--477, 2018.

\bibitem{sas2022evolution}
\BIBentryALTinterwordspacing
D.~Sas, P.~Avgeriou, and U.~Uyumaz, ``On the evolution and impact of
  architectural smells—an industrial case study,'' \emph{Empirical Software
  Engineering}, vol.~27, no.~86, 2022. [Online]. Available:
  \url{https://doi.org/10.1007/s10664-022-10132-7}
\BIBentrySTDinterwordspacing

\bibitem{designite2016}
T.~Sharma, P.~Mishra, and R.~Tiwari, ``Designite: A software design quality
  assessment tool,'' in \emph{International Workshop on Bringing Architectural
  Design Thinking into Developers’ Daily Activities}, 2016, pp. 1--4,
  international Workshop on Bringing Architectural Design Thinking into
  Developers’ Daily Activities.

\bibitem{singh2018}
S.~Singh and S.~Kaur, ``A systematic literature review: Refactoring for
  disclosing code smells in object oriented software,'' \emph{Ain Shams
  Engineering Journal}, vol.~9, no.~4, pp. 2129--2151, 2018.

\bibitem{7}
\BIBentryALTinterwordspacing
A.~Vaswani, N.~Shazeer, N.~Parmar, J.~Uszkoreit, L.~Jones, A.~N. Gomez,
  L.~Kaiser, and I.~Polosukhin, ``Attention is all you need,'' in
  \emph{Advances in Neural Information Processing Systems}, vol.~30.\hskip 1em
  plus 0.5em minus 0.4em\relax Curran Associates, Inc., 2017. [Online].
  Available:
  \url{https://proceedings.neurips.cc/paper_files/paper/2017/file/3f5ee243547dee91fbd053c1c4a845aa-Paper.pdf}
\BIBentrySTDinterwordspacing

\bibitem{wang2021codet5}
Y.~Wang, W.~Wang, S.~Joty, and S.~C. Hoi, ``Codet5: Identifier-aware unified
  pre-trained encoder-decoder models for code understanding and generation,''
  \url{https://arxiv.org/abs/2109.00859}, 2021, arXiv preprint
  arXiv:2109.00859.

\bibitem{XIAO2025107067}
\BIBentryALTinterwordspacing
Y.~Xiao, X.~Zuo, X.~Lu, J.~S. Dong, X.~Cao, and I.~Beschastnikh, ``Promises and
  perils of using transformer-based models for se research,'' \emph{Neural
  Networks}, vol. 184, p. 107067, 2025. [Online]. Available:
  \url{https://www.sciencedirect.com/science/article/pii/S0893608024009961}
\BIBentrySTDinterwordspacing

\bibitem{Zakeri_Nasrabadi_2023}
\BIBentryALTinterwordspacing
M.~Zakeri-Nasrabadi, S.~Parsa, E.~Esmaili, and F.~Palomba, ``A systematic
  literature review on the code smells datasets and validation mechanisms,''
  \emph{ACM Computing Surveys}, vol.~55, no. 13s, p. 1–48, Jul. 2023.
  [Online]. Available: \url{http://dx.doi.org/10.1145/3596908}
\BIBentrySTDinterwordspacing

\bibitem{scsmell2024}
D.~Zhang, S.~Song, Y.~Zhang, and H.~Liu, ``Code smell detection research based
  on pre-training and stacking models,'' \emph{IEEE Latin America
  Transactions}, vol.~22, no.~1, pp. 22--30, 2024.

\bibitem{zhang2011}
M.~Zhang, T.~Hall, and N.~Baddoo, ``Code bad smells: A review of current
  knowledge,'' \emph{Journal of Software Maintenance and Evolution: Research
  and Practice}, vol.~23, no.~3, pp. 179--202, 2011.

\bibitem{zhang2024}
Y.~Zhang, C.~Ge, H.~Liu, and K.~Zheng, ``Code smell detection based on
  supervised learning models: A survey,'' \emph{Neurocomputing}, vol. 565, p.
  127014, 2024, dOI:10.1016/j.neucom.2023.127014.

\end{thebibliography}

\end{document}